\documentclass[aps, showpacs, twocolumn]{revtex4}

\usepackage{amsmath,amssymb,amsfonts}
\usepackage{tabularx}
\usepackage{epsfig}
\usepackage{graphicx}

\renewcommand{\L}{{\cal L}}
\renewcommand{\lambdabar}{{\overline\lambda}}

\newcommand{\psibar}{{\overline\psi}}

\newcommand{\alphadot}{{\dot\alpha}}

\def\pslash#1{{\setbox0=\hbox{$#1$}
  \rlap{\ifdim\wd0>.7em\kern.22\wd0\else\kern.1\wd0\fi /}#1}}

\begin{document}

\title{Critical behavior of a supersymmetric extension of the Ginzburg-Landau model}

\author{Grigoris Panotopoulos\footnote{Grigoris.Panotopoulos@uv.es}}

\date{\today}


\address{Departament de Fisica Teorica, Universitat de Valencia, E-46100 Burjassot, Spain, and
Instituto de Fisica Corpuscular (IFIC), Universitat de 
Valencia-CSIC, Edificio de Institutos de Paterna, Apt. 22085, E-46071, 
Valencia, Spain.}


\begin{abstract}
We make a connection between quantum phase transitions in condensed matter
systems, and supersymmetric gauge theories that are of interest in the particle
physics literature. In particular, we point out interesting effects of the
supersymmetric quantum electrodynamics upon the critical behavior of the
Ginzburg-Landau model. It is shown that supersymmetry fixes the critical exponents, 
as well as the Landau-Ginzburg parameter, and that the model resides in the type
II regime of superconductivity.
\end{abstract}


\maketitle

\section{Introduction}

A very well studied model in the condensed matter literature is the Ginzburg-Landau (GL) 
model~\cite{LG},
described by the lagrangian of an Abelian Higgs model
\begin{equation}
\mathcal{L}=|D_\mu \phi|^2+m^2|\phi|^2+\lambda|\phi|^4+\frac{1}{4}F_{\mu \nu}^2
\end{equation}
where $\phi$ is a complex scalar field charged under the abelian gauge field
$A_\mu$, with the gauge covariant derivative and field strength
\begin{eqnarray}
D_\mu & = & \partial_\mu-ie A_\mu\ \\
F_{\mu\nu} & = & \partial_\mu A_\nu - \partial_\nu A_\mu\ 
\end{eqnarray}
When $m^2>0$ the gauge symmetry is exact, and the model describes a
massive complex scalar particle that interacts with a massless photon. The
electric potential between these scalars has the usual Coulomb form, and
therefore this is referred to as the Coulomb phase. On the other hand, when
$m^2<0$ the gauge symmetry is spontaneously broken, and in this Higgs phase
the model describes a massive gauge boson and a massive real scalar field.
The nature of the transition between the Higgs and Coulomb phase has been
of great interest to the condensed matter community. 

The critical fluctuations
in the Ginzburg-Landau model of superconductors were studied in~\cite{old1}, while
the fixed point structure for the GL model was presented in~\cite{old2}.
Furthermore, in previous works the authors have investigated models in which massless 
Dirac fermions 
are coupled to the Ginzburg-Landau model~\cite{nogueira}. The presence of the 
Dirac fermions is justified 
by the fact
that effective microscopic models of strongly correlated electrons usualy
contain them~\cite{fermions}. The critical exponents can be computed as a function of 
the number $N_F$
of the fermions, and for increasing $N_F$ the models is driven into the type II regime
of superconductivity. In particular, for the minimum allowed value of the fermion number,
$N_F=4$, both values of the $\kappa$ parameter, corresponding to the 'T' fixed point and
the 'SC' fixed point, are found to be above the mean-field GL value $1/\sqrt{2}$, in 
contrast to the theoretical~\cite{theory} and the Monte Carlo numbers~\cite{MC} in
the GL model. In this article we point out that the generalization of the 
Ginzburg-Landau model to a supersymmetric one necessarily introduces fermions both in the 
matter and gauge supermultiplets, and that the restrictions imposed by the symmetries of 
the model unambigiously determine the critical exponents and the Landau-Ginzburg parameter, 
which is found to be in the type II regime of superconductivity.

Finally, we remind the reader that a) all
exactly solvable models show that not all of the critical exponents
are independent. In fact they satisfy certain scaling laws, supported by
all the experimental and
numerical results, and it can be shown that
there are only two independent critical exponents. If we take them to be $\eta$ and
$\nu$, the rest of the critical exponents are given by~\cite{bellac, kleinert}
\begin{eqnarray}
\alpha & = & 2-\nu D \\
\beta & = & \frac{\nu}{2} (D-2+\eta) \\
\gamma & = & \nu (2-\eta) \\
\delta & = & \frac{D+2-\eta}{D-2+\eta}
\end{eqnarray}
where $D$ is the dimension of the system, and b)
in the Landau-Ginzburg
theory there are two fundamental length scales, namely the
penetration length $\lambda$ and the coherence length $\xi_0$. The
Landau-Ginzburg parameter $\kappa$ is defined as follows
\begin{equation}
\kappa \equiv \frac{\lambda}{\xi_0}
\end{equation}
and it can be shown that $\kappa < 1/\sqrt{2}$ corresponds to
type I superconductors, while $\kappa > 1/\sqrt{2}$ corresponds to
type II superconductors.

\section{The supersymmetric model and critical exponents}

Supersymmetric Quantum Electrodynamics (SQED) is an abelian gauge theory
with the following field content:~\cite{sqed}
\begin{enumerate}
\item One vector multiplet $(A^\mu, \lambda^\alpha,
  \lambdabar^\alphadot)$ consisting of the
  photon and the photino (in the so-called Wess-Zumino gauge), described 
by a vector and a Majorana spinor field.
\item Two chiral multiplets $(\psi_L^\alpha, \phi_L)$ and
  $(\psi_R^\alpha, \phi_R)$ with charges $Q_L = -1$, $Q_R = +1$, each
  consisting of one Weyl spinor and one scalar field, constituting
  the left- and right-handed electron and selectron, the matter
  fields.
\end{enumerate}
The electron Dirac spinor and the photino Majorana spinor are given by
\begin{eqnarray}
\Psi = {{\psi_L}_\alpha \choose \psibar_R^\alphadot} ,\quad
\tilde{\gamma} = {-i\lambda_\alpha \choose i\lambdabar^\alphadot}\ .
\label{4Spinors}
\end{eqnarray}
The SQED Lagrangian contains kinetic, minimal coupling and mass terms
and in addition, due to the supersymmetry, coupling terms to the
photino and quartic terms in the selectron fields:
\begin{eqnarray}
{\L}_{\rm SQED} & = &
-\frac{1}{4}F_{\mu\nu}F^{\mu\nu} +
\frac{1}{2}\overline{\tilde{\gamma}}i\gamma^\mu\partial_\mu\tilde{\gamma}
\nonumber\\
&&{}
+|D_\mu \phi_L|^2 + |D_\mu \phi_R^\dagger|^2
+ \overline{\Psi}i\gamma^\mu D_\mu \Psi
\nonumber\\
&&{}
-\sqrt{2}eQ_L\left(
\overline{\Psi}P_R\tilde{\gamma}\phi_L
- \overline{\Psi}P_L\tilde{\gamma} \phi_R^\dagger
\right.
\nonumber\\
&&
\phantom{-\sqrt{2}eQ_L(}
\left.
+ \phi_L^\dagger \overline{\tilde{\gamma}}P_L \Psi
- \phi_R \overline{\tilde{\gamma}}P_R \Psi \right)
\nonumber\\
&&{}
-\frac{1}{2}\left(eQ_L|\phi_L|^2 + eQ_R|\phi_R|^2\right)^2
\nonumber\\
&&{}
-m\overline{\Psi} \Psi - m^2 (|\phi_L|^2 + |\phi_R|^2)
\label{LSQED}
\end{eqnarray}
with the gauge covariant derivative and field strength
\begin{eqnarray}
D_\mu & = & \partial_\mu+ieQ A_\mu\ ,\\
F_{\mu\nu} & = & \partial_\mu A_\nu - \partial_\nu A_\mu\ .
\end{eqnarray}
It must be noted that the model with just one chiral supermultiplet
is anomalous, while the inclusion of a second chiral supermultiplet
with opposite electric charge renders the model anomaly-free, since
in this case $TrQ=0$. The two Weyl spinors combine to form the Dirac
spinor of the usual spinor electrodynamics in the standard
four-component formalism.

The model contains both bosons and fermions, with equal masses and
degrees of freedom within each multiplet. The form of the
interactions, as well as the values of the couplings, are completely
determined by the symmetries. It is interesting that there is just
one coupling constant, namely the electric charge $e$. We have the
usual types of interaction that one encounters in the usual field
theory, namely quartic interaction for the scalars, Yukawa coupling,
and the gauge (electromagnetic) interaction. We thus know that the
theory is renormalizable. In fact, here we have just a wave function
renormalization both for the vector and the chiral multiplets due to
supersymmetry~\cite{theorem}, and furthermore the beta function for
the electric charge is determined by the photon self-energy and
wave-function renormalization due to gauge invariance~\cite{BD}.

The investigation regarding the critical behavior is according to
the following program: a) Perform a one-loop analysis to compute the
relevant counterterms that eliminate the unwanted divergencies, b)
determine the beta-function for the electric charge $\beta(e)$, as
well as the anomalous dimensions for the scalars $\gamma_m,
\gamma_{\phi}$, c) find the fixed points from the condition
$\beta(e_*)=0$, and finally d) compute the critical exponents $\eta,
\nu$ using the well-known formulas~\cite{nogueira, nogueira2}
\begin{eqnarray}
\eta & = & 2 {\gamma_{\phi}}_* \\
\nu  & = & \frac{1}{2 (1-{\gamma_m}_*)}
\end{eqnarray}
where the anomalous dimensions, as well as the beta function are 
given by~\cite{fieldtheory}
\begin{eqnarray}
\beta(\alpha) & = &  \mu \: \frac{\partial{\alpha}}
{\partial{\mu}} \\
\gamma_{\phi} & = & \frac{1}{2} \: \mu \: \frac{\partial{\textrm{ln} Z_{\phi}}}
{\partial{\mu}} \\
\gamma_m  & = & \frac{\mu}{m} \: \frac{\partial{m}}
{\partial{\mu}}
\end{eqnarray}
with $\mu$ the renormalization mass scale, and $\alpha=e^2/4 \pi$ the 
fine-structure constant. Note that our definitions for the anomalous dimensions
are slightly different than~\cite{nogueira, nogueira2}.

\begin{figure}
\centering
\includegraphics[width=\linewidth]{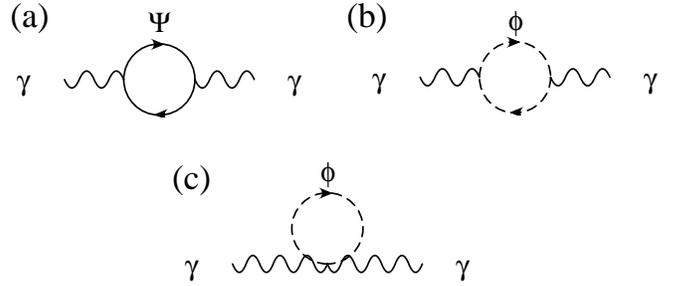}
\caption{Feynman diagrams for the photon self-energy with
the usual spinor and scalar electrodynamics
interaction vertices.}
\end{figure}

We start from the photon self energy, that will allows to compute 
the electric charge beta function $\beta(\alpha)$. The relevant loop-diagrams are
shown in Figure~1. We have the same diagrams as in the usual spinor
and scalar electrodynamics. The electric charge beta function has a contribution
from a Dirac spinor and a contribution from two complex scalars. At one loop and
using dimensional regularization~\cite{'tHooft:1972fi} 
(the space-time dimension 4 $\rightarrow$ $D=4-\epsilon$,
then take the limit $\epsilon \rightarrow 0$ and isolate the divergent
part $\sim 1/\epsilon$) one obtains the result
\begin{equation}
\beta(\alpha, \epsilon) =-\epsilon \alpha+\frac{\alpha^2}{\pi}
\end{equation}

\begin{figure}
\centering
\includegraphics[width=\linewidth]{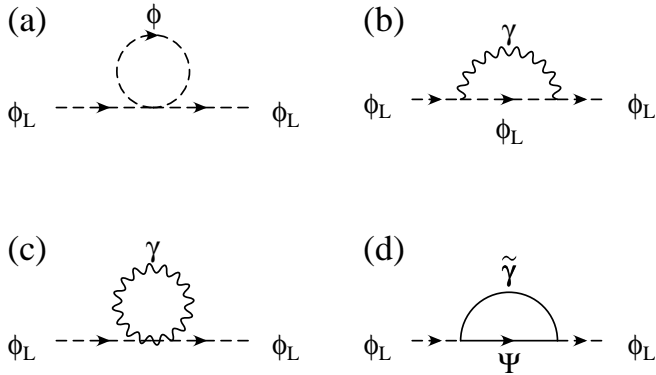}
\caption{Feynman diagrams for the scalar self-energy with 
(a) the quartic,
(b) the single photon,
(c) the two-photon, and
(d) the Yukawa interaction vertices.}
\end{figure}

Next we turn to the scalar field self-energy. The relevant diagrams can be
shown in Figure~2. We have the three usual diagrams from scalar
electrodynamics, plus a new one with the Yukawa coupling with the Dirac electron
and the photino Majorana fermion. For the scalar field wave-function
renormalization we find the result (in the Lorentz gauge)
\begin{equation}
Z_{\phi} = 1+\frac{5 e^2}{8 \pi^2 \epsilon}
\end{equation}
Now it is a straightforward algebraic task to compute the anomalous
dimensions and then the critical exponents. We thus obtain our final
results (for $D=3$ or $\epsilon=1$)
\begin{eqnarray}
\eta & = & -2.5 \\
\nu & = & \frac{1}{6} \simeq 0.17
\end{eqnarray}
Our results for $\beta(e)$ and $\gamma_{\phi}$ agree with the corresponding 
formulas of~\cite{Machacek:1983tz} at one loop.
In~\cite{nogueira} there are $N_F$ massless fermions, and two
coupling constants with two different beta functions. The authors
in~\cite{nogueira} have found two fixed points (tricritical and 
superconducting), and that the number
of massless fermions must be at least four. The $\eta$ critical
exponent is always negative, while the $\nu$ critical exponent is
always positive, and for both exponents the absolute value is a
number around $0.5$ when $N_F$ is small. In our supersymmetric version of the model,
there is just one massive fermion, since supersymmetry requires that
there are equal number of fermionic and bosonic degrees of freedom,
and with the same masses. There is only one coupling constant,
namely the electric charge $e$, and thus just one beta function, and
a single infrared stable fixed point. Despite this, there is also here a quartic
self-interaction potential for the scalar field, where the coupling
is fixed by supersymmetry, and it is given in terms of the electric
charge. Furthermore, we find also a negative $\eta$ critical
exponent and a positive $\nu$ exponent, with values not too different 
from the ones obtained in~\cite{nogueira} for small $N_F$.

\section{Supersymmetry breaking and the $\kappa$ parameter}

So far we have not seen any superpartners yet, and thus supersymmetry must
be broken. In this section we shall discuss spontaneous breaking of
supersymmetry, following~\cite{wessbagger}, within the framework of the 
Fayet-Iliopoulos mechanism~\cite{Fayet:1974jb}.
In an abelian $U(1)$
supersymmetric
gauge theory an extra term is allowed by the symmetries, the so called
Fayet-Iliopoulos term, $\xi D$, where $D$ is the auxiliary field in the vector
supermultiplet,
and $\xi$ is a new parameter with mass dimension two. If $F_1, F_2, D$ are
the auxiliary fields in the off-shell formulation of the supersymmetric theory,
the scalar potential is given by
\begin{equation}
\mathcal{V} = \frac{1}{2} D^2+F_1 F_1^*+F_2 F_2^*
\end{equation}
and the auxiliary fields satisfy the following equations of motion
\begin{eqnarray}
F_1+m A_2^* & = & 0 \\
F_2+m A_1^* & = & 0 \\
D+\xi+\frac{e}{2} (A_1 A_1^*-A_2 A_2^*) & = & 0
\end{eqnarray}
where now the scalar fields are denoted by $A_1, A_2$ instead of
$\phi_L, \phi_R$. Supersymmetry is broken since there is no solution
that leaves $\mathcal{V}=0$. Upon substitution the scalar potential
takes the form
\begin{eqnarray}
\mathcal{V}&=&\frac{\xi}{2}+(m^2+\frac{e \kappa}{2}) A_1 A_1^*+
(m^2-\frac{e \xi}{2}) A_2 A_2^*
\nonumber\\
&&
+\frac{1}{8}e^2 ( A_1 A_1^*- A_2 A_2^*)^2
\end{eqnarray}
We can see that there are two possibilities, namely that $m^2 > e
\xi/2$ or $m^2 < e \xi/2$. In the first case the $A_1=0=A_2$
minimizes the potential, the form of which is shown in Figure~3(a).
The supersymmetry is spontaneously broken but the gauge symmetry is
exact. The theory describes two complex scalar fields with masses
$m^2+\frac{e \xi}{2}$ and  $m^2-\frac{e \xi}{2}$. The rest of the
fields, namely the photon $\gamma$, the photino $\lambda$, and the
two fermions $\psi_1, \psi_2$ retain their masses. In particular,
the photino is the massless goldstino. In the second case the
$A_1=0=A_2$ no longer minimizes the potential, the form of which is
shown in Figure~3(b). This time both the supersymmetry and the gauge
symmetry are broken simultaneously. The minimum corresponds to
$A_1=0, A_2=v$, where the vacuum expectation value $v$ is determined
from
\begin{equation}
\frac{e^2 v^2}{4}+(m^2-\frac{e \xi}{2})=0
\end{equation}
This model describes a vector field and a scalar field of mass
$\sqrt{\frac{e \xi}{2}}$, a complex scalar field with mass
$\sqrt{2 m^2}$, a massless goldstino, and two spinor fields with mass
$\sqrt{m^2+\frac{e \xi}{2}}$. The Landau-Ginzburg parameter therefore
is easily computed to be
\begin{equation}
\kappa=\frac{m_s}{m_v}=1 \simeq \frac{1.41}{\sqrt{2}}
\end{equation}
which is larger than $1/\sqrt{2}$, and we thus have a type II 
superconductor. It is interesting to see again that our value of
the Ginzburg parameter is comparable to the value obtained 
in~\cite{nogueira} at the superconducting fixed point and for $N_F=4$.
Therefore, we conclude that supersymmetry provides the kind of lagrangian 
studied in~\cite{nogueira}, and that the values of the Ginzburg parameter
and of the critical exponents are similar to the ones obtained 
in~\cite{nogueira}, without a second coupling constant $\lambda$ for the scalar
quartic self-interaction, and without many fermions.

\begin{figure}
\centering
\includegraphics[width=\linewidth]{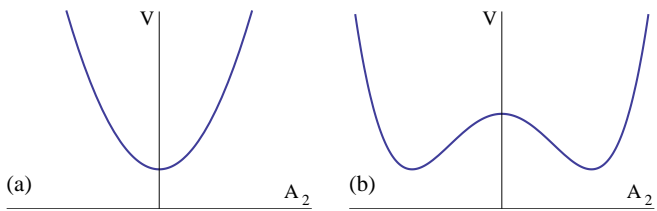}
\caption{(a) The scalar potential versus $A_2$ in the $m^2 > e \kappa/2$ case (in arbitrary
units). Supersymmetry is spontaneously broken, but the $U(1)$ gauge symmetry
is exact.
(b) As in (a) but in the $m^2 < e \kappa/2$ case.
Here, both supersymmetry and $U(1)$ gauge symmetry are spontaneously
broken.}
\end{figure}


\section{Conclusions}

We have proposed and analyzed a supersymmetric extension of the
Landau-Ginzburg theory, which is essentially the supersymmetric
version of quantum electrodynamics. The model describes the interaction of a
Dirac fermion and two complex scalar fields with the photon and its
superpartner, the photino, which is a Majorana fermion. All the
couplings in the model are given in terms of the electric charge. It
is interesting that there is a quartic self-interaction coupling for 
the scalar fields
even in the absence of a coupling $\lambda$. Within the
one-loop renormalization program we give the expression for the
wave-function renormalization, and according to the standard prescription we
compute the critical exponents $\eta, \nu$ from the beta function
and the anomalous dimensions. Finally, we have discussed spontaneous
supersymmetry breaking a la Fayet-Iliopoulos mechanism. There is a
case in which both supersymmetry and gauge symmetry can be broken at
the same time. The photon acquires a non-vanishing mass, and the
Landau-Ginzburg parameter is computed. We find that its value
corresponds to type II superconductors. Our values of the Ginzburg
parameter and of the critical exponents are similar to the ones 
obtained in~\cite{nogueira}, without many fermions and without the 
introduction of a second coupling constant for the scalar 
quartic self-interaction.

\section*{Acknowledgments}

The author acknowledges financial
support from FPA2008-02878 and Generalitat Valenciana under the grant PROMETEO/2008/004.


\end{document}